\documentclass[a4paper,12pt]{article}

\usepackage[latin1]{inputenc}
\usepackage[top=2.5cm, left=2.5cm, right=2cm, bottom=2.5cm]{geometry}
\usepackage{graphicx}
\usepackage[centertags]{amsmath}
\usepackage{amsfonts}
\usepackage{amssymb}
\usepackage{amsthm}
\usepackage{newlfont}

\newcommand{\beq}{\begin{equation}}
\newcommand{\eeq}{\end{equation}}
\newcommand{\bea}{\begin{eqnarray}}
\newcommand{\eea}{\end{eqnarray}}
\newcommand{\p}{\partial}

\begin{document}

\begin{center}

{\Large \bf{Hamilton-Jacobi analysis of the four dimensional $BF$ model with cosmological term}}

\vspace{1cm}

G. B. de Gracia \footnote{gb9950@gmail.com}, B. M. Pimentel\footnote{pimentel@ift.unesp.br}, C. E. Valc\'arcel\footnote{valcarcel.flores@gmail.com}

\vspace{.5cm}

$^{1,2}$\emph{Instituto de F\'isica Te\'orica, UNESP - S\~ao Paulo State University},

\emph{P. O. Box 70532-2, 01156-970, S\~ao Paulo, SP, Brazil}.

$^{3}$\emph{Centro de Matem\'atica, Computa\c{c}\~ao e Cogni\c{c}\~ao},

\emph{Universidade Federal do ABC, 09210-170 Santo Andr\'e, SP, Brazil}.

\end{center}

\vspace{.25cm}

\begin{abstract}

In this work we perform the Hamilton-Jacobi constraint analysis of the four dimensional Background Field ($BF$) model with cosmological term. We obtain the complete set of involutive Hamiltonians that guarantee the integrability of the system and identify the reduced phase space. From the fundamental differential we recover the equations of motion and obtain the generators of the gauge and shift transformations.

\vspace{.5cm}


\end{abstract}

\section{Introduction}

The importance of the Background Field ($BF$) theories lies on the fact that they have a close relationship with gravity $\cite{F1}$. Those theories are topological theories and they do not depend on the space-time metric along with its correlation functions and, at principle, its quantization would be easier than the one of the Einsten-Hilbert action, for example. There is plenty of literature related to the application of techniques such as spin foam quantization to $BF$ theories as a method to have some insights about the quantum behavior of gravity $\cite{Ba2}$, $\cite{Per}$.

Lower dimensional $BF$ models are also good laboratories to analyze lower dimensional quantum gravity. Those models have a common feature: all of them propagates zero degrees of freedom. This fact is in accordance with its gravitational interpretation since gravity in lower dimensions have no degrees of freedom.

In four dimensions, there are two important $BF$ models of gravity: The Plebanski theory \cite{Pleb} and the Freidel-Starodubtsev model \cite{F2}, in both cases they begin with a four-dimensional $BF$ model plus a cosmological term. In the Plebanski theory it is introduced an additional field such that it is imposed simplicity constraints. The Freidel-Starodubtsev model is equivalent to the MacDowell-Mansouri gravity \cite{Mc}, this is shown by introducing an interaction term which breaks the original symmetry of the $BF$ model. Recently, it has been shown that a similar construction of the Freidel-Starodubsetv can be applied for two and three-dimensional gravity \cite{Paszko} with a Polynomial $BF$ action. For a recent review on $BF$ gravity see \cite{Montesinos}.

In order to identify the true degrees of freedom and dynamical variables of the $BF$ models, it is important to analyze the constraint structure. Usually, this procedure is made with the Hamiltonian Dirac formalism \cite{Dirac} (also see \cite{const_books}). The Hamiltonian analysis of Plebanski theory has been made in \cite{HamPle}, the Freidel-Starodubtsev in \cite{BFHam}, the two-dimensions Polynomial $BF$ in \cite{Valcarcel} and the $BF$ with cosmological term in \cite{Esc}. However, there are other methods of constraint analysis, as the Hamilton-Jacobi formalism.

The Hamilton-Jacobi formalism presented here follows the approach of G\"{u}ler \cite{Gul}, which is an extension of Caratheodory's equivalent Lagrangian method in the calculus of variations \cite{Carat}. This formalism is characterized by a set of Hamilton-Jacobi differential equations called Hamiltonians. The dynamical evolution of the system is given in terms of a fundamental differential which depends on the time and other linear independent arbitrary parameters related to the involutive Hamiltonians \cite{Non}, \cite{Inv}, obtained from the Frobenius' Integrability Condition. The canonical transformations are obtained immediately from this fundamental differential when just the dynamics described by those arbitrary parameters are considered. Furthermore, the gauge transformations are the subgroup of those transformations that leave the lagrangian invariant. On the other hand, although the Dirac's approach is a very powerful tool to the constraint analysis, it deals with gauge symmetries by the conjecture that they are all generated by the first class constraints of the theory. Unfortunately, there are some examples that contradict it \cite{Git}. Therefore, we claim again that the Hamilton-Jacobi formalism is a way to illuminate the canonical origin of the gauge structure of the four dimensional $BF$ theory. This approach was used to study several examples of gauge systems such as topologically massive theories \cite{Topol}, gravity models \cite{Grav}, and the two dimensional and three dimensional $BF$ theories \cite{BF2}, \cite{BF3}. This formalism were also extended to higher order Lagrangians and Berezinian systems \cite{Ber}.

In this work, we will deal with the constraint analysis of the four dimensional $BF$ theory with a cosmological term. This model is a natural extension of the lower dimensional ones in $1+1$ \cite{BF2} and in $2+1$ \cite{BF3} dimensions studied under the Hamilton-Jacobi formalism. A careful analysis of its constraint structure and symmetry properties is an excellent laboratory to a future investigation of the gravitational $BF$ models since it can give us insights about the implications of adding a symmetry breaking term in its highly symmetric structure. We argue that a Hamilton-Jacobi analysis can be enriching because it deals with gauge symmetries in a very natural way.

The following section will be devoted to explain the main features of the Hamilton-Jacobi formalism. In section $3$ we present some general characteristics of the four dimensional $BF$ model. In section $4$ we give its constraint analysis and build the generalized brackets. In the section $5$ we compute its characteristic equations generated by the fundamental differential and analyze the equivalence with the Lagrangian formalism, as well as obtain the canonical and gauge transformations, and from them we find generators of those  transformations. Finally, in section $6$ we discuss the results.

\section{The Hamilton-Jacobi formalism}

In this section we develop the Hamilton-Jacobi formalism for constrained systems, which are defined as the  ones  whose Lagrangian do not satisfy the Hessian condition.

Let us consider a physical system whose Lagrangian has the form $L=L(x^i,\ \dot x^i,\ t)$ where the Latin indices run from $1$ to $n$, which is the dimension of the configuration space. The system is called constrained or singular if it does not satisfy the Hessian condition $\det W_{ij}\neq 0$ with the matrix $W_{ij}$ given by $W_{ij}=\frac{\p^2L}{\p\dot x^i  \p \dot x^j}$. If the Hessian condition is satisfied, the transformation that leads the configuration space to the phase space is invertible. If it is not, some of the conjugated momenta $p_i=\frac{\p L}{\p \dot x^i}$ are not invertible on velocities and we are lead to equations of the form $\Phi(q,p)=0$ which constrains the phase space. Now, if we consider $k$ non-invertible momenta and $m=n-k$ invertible momenta, we have
\bea
p_z-\frac{\p L}{\p \dot x^z}=0,
\eea
where $z=1,...k$. Then, the above equation defines the primary constraints of the theory. By using the definition $H_z\equiv -\frac{\p L}{\p \dot x^z}$ we can rewrite the above equation as
\bea
H_z'\equiv p_z+H_z=0.
\eea
These constraints are called Hamiltonians. If we define $p_0\equiv \frac{\p S}{\p t}$, the Hamilton-Jacobi equation can be written as
\bea
H_0'\equiv p_0+H_0=0.
\eea

The canonical Hamiltonian function $H_0=p_a\dot x^a+p_z\dot x^z-L$ with $a=1,...,m$, is independent of the non-invertible velocities $\dot x^z$ if the constraints are implemented. The unified notation is given by
\bea
H_\alpha'\equiv p_\alpha+H_\alpha,
\eea
where $\alpha=0,...,k.$. The Cauchy's method \cite{Carat} is employed to find the characteristic equations related to the above first order equations
\bea
dx^a=\frac{\p H_\alpha'}{\p p_a}dt^\alpha, \quad  \quad dp^a=-\frac{\p H_\alpha'}{\p x_a}dt^\alpha, \quad  \quad dS=(p_adx^a-H_\alpha dt^\alpha).
\eea
The differentials written above depend on $t^\alpha=(t^0,t^z\equiv x^z)$ independent variables or parameters. The name Hamiltonians used for the constraints is now justified, once that $H_z$ generates flows parameterized by $t^z$ in analogy with the temporal evolution generated by $H_0$. From the characteristic equations one can use the Poisson brackets defined on the extended phase space $(x^a,t^\alpha,p_a,p_\alpha)$ to express in a concise form the evolution of any function $f=f(x^a,t^\alpha,p_a,p_\alpha)$:
\bea
df=\{f,H'_\alpha\}dt^\alpha. \label{IC01}
\eea
This is the fundamental differential, from where we identify the Hamiltonians as the generators of the dynamical evolution of the phase-space functions.

Let us define the operator
\bea
X_\alpha[f]=\sum_I\{\gamma^I,H'_\alpha\}\frac{\delta f}{\delta \gamma^I}\quad ; \quad \gamma^I=(x^a,t^\alpha,p_a,p_\alpha),\label{IC02}
\eea
where $X_\alpha[*]$ can be interpreted as $k$ vectors whose $2(n+1)$ components are $\{\gamma^I,H'_\alpha\}$. The fundamental differential can be expressed in terms of this operator as
\bea
df = X_\alpha[f] dt^\alpha.\label{IC03}                                                                                                                                         \eea
The Frobenius' Integrability Condition (IC) ensures the system of equations (\ref{IC03}) is integrable. The IC can be expressed as
\bea
\big[X_\alpha,X_\beta \big](x^a,p^a)\equiv  X_\alpha [x^a]  X_\beta [p^a]-X_\beta[x^a]  X_\alpha[p^a]=-\{H'_\alpha, H'_\beta \}=0.  \label{IC04}
\eea
The above condition can be generalized to
\bea
\{H'_\alpha, H'_\beta \}=C^\gamma_{\alpha \beta}H_\gamma', \label{IC05}
\eea
where $C^\gamma_{\alpha \beta}$ are structure coefficients (See \cite{Mishchenko}). Therefore, the IC ensures that the Hamiltonians close an involutive algebra. In terms of the fundamental differential (\ref{IC01}), the IC (\label{IC05}) can be written as
\bea
dH'_\alpha =\{H'_\alpha,H'_\beta\}dt^\beta=C^\gamma_{\alpha \beta}H_\gamma'dt^\beta = 0^\beta. \label{IC06}
\eea
The Hamiltonians that satisfy the IC are called involutives. However, not all Hamiltonians from a  physical systems satisfy this condition identically. Therefore, we must define new Hamiltonians.

Let us suppose we have a set of non-involutive Hamiltonians $H'_{\bar a}$. Then
\bea
dH'_{\bar a} = \{H'_{\bar a},H'_0\} dt + \{H'_{\bar a},H'_{\bar b}\}dx^{\bar b}.\label{IC06c}
\eea
Once that we impose $dH'_{\bar a}=0$, we can define a matrix with components $M_{\bar{a} \bar{b}} \equiv \{H'_{\bar a},H'_{\bar a}\}$. If this matrix is invertible, we can write $dx^b = -M^{-1}_{\bar{a}\bar{b}} \{H'_{\bar a},H'_0\} dt$, i.e., there is a dependence between the parameters related to the non-involutive Hamiltonians. Replacing in the fundamental differential, we have
\bea
dF = [\{F,H'_0\} - \{F,H'_{\bar a}\}M^{-1}_{\bar{a} \bar{b}} \{H'_{\bar b},H'_0\}] dt.\label{IC06d}
\eea
Therefore, we can define Generalized Brackets (GB) as:
\bea
\{A,B\}^*\equiv \{A,B\}-\{A,H'_{\bar a}\}({M}^{-1})_{\bar a \bar b}\{H'_{\bar a},B\},\label{IC07}
\eea
which redefine the dynamic of the constrained system reducing its phase-space, once that $dF=\{F,H'_0\}^*dt$. This procedure is the result of the Integrability Condition and, as shown in \cite{Non}, it allows the possibility that the matrix $M_{ab}$ is non-invertible, or that the system has involutive and non-involutive Hamiltonians as well.

The dynamical evolution described by the resulting arbitrary parameters can be understood as canonical transformations, with the involutive Hamiltonians as generators. To understand this, we need to check that the variation $\delta\gamma^I = \delta t^\alpha X_\alpha[\gamma^I]$ is generated by $g=1+\delta t^\alpha X_\alpha$, also preserves the symplectic structure $dx^a\wedge dp_a+dt^\alpha\wedge dp_\alpha+dH_\alpha\wedge dt^\alpha$. with fixed $dt^0$. In order to relate canonical transformations with the gauge ones, we need to restrict the study to fixed times $dt^0 = 0$. Then, the transformation on any variable $\gamma^I$ is
\bea
\gamma^I=\{\gamma^I, H'_z\}^*\delta t^z.\label{IC08}
\eea
The Hamiltonians must be involutives, then $\{H'_x, H'_y\}^*=C^z_{xy}H'_z$. However, the IC ensures that $\{H'_x, H'_y\}^*=C^0_{xy}H'_0+C^z_{xy}H'_z$. To conciliate these equations we must consider whether $C^0_{xy}=0$ or $H'_0=0$. The condition $C^0_{xy}=0$ is almost never satisfied. On the other hand, the condition $H'_0=0$ constrains the phase-space. Under this assumption, we define the generator of gauge transformations as
\bea
G^{can} \equiv H'_z \delta t^z,\label{IC09}
\eea
since $\delta \gamma^I =\{\gamma^I, G^{can}\}^* $. More details on the role of involutive Hamiltonians in the HJ formalism can be found in \cite{Inv}.

\section{The four-dimensional $BF$ model with Cosmological Term}

Let us consider the gauge group $SO(1,3)$ acting on a background field $2-$form $B$ and a connection $1-$form $A$ of a four-dimensional manifold $\mathcal{M}$. The generators of the gauge group $M_{IJ}=-M_{JI}$ obey the following algebra:
\bea
[M_{IJ},M_{KL}]=\eta_{IL}M_{JK}-\eta_{IK}M_{JL}+\eta_{JK}M_{IL}-\eta_{JL}M_{IK},\label{BF01}
\eea
where $I,J=0,1,2,3$ and $\eta_{IJ}=diag(+,-,-,-)$. We can also define the $2-$form strength field as $F=dA + A\wedge A$.

The action for the $BF$ model is characterized by $tr(B\wedge F)$, which is gauge invariant due to the properties of the trace and wedge product. In four dimensions we can also add the expression $tr(B\wedge B)$, called cosmological term, which also maintains the invariance properties of the $BF$ model. The action of the $BF$ model with cosmological term is given by
\bea
S=\int_{\mathcal M}\, tr \bigg (B\wedge F-\frac{\beta}{2} B\wedge B\bigg ),\label{BF02}
\eea
where $\beta$ is a constant parameter.

Once, the $B$ field and connection $A$ acts on the $SO(1,3)$, we can write
\bea
A=A_\mu^{IJ} M_{IJ}dx^\mu,\quad B=\frac{1}{2}B_{\mu \nu}^{IJ} M_{IJ}dx^\mu \wedge dx^\nu.\label{BF03}
\eea
Furthermore, from the definition of field strength, we obtain
\bea
F_{\mu\nu}^{IJ} = \p_\mu A^{IJ}_\nu-\p_\nu A_{\mu}^{IJ}+A_{\mu K}^IA_\nu^{KJ}-
A_{\nu K}^{I}A_{\mu}^{KJ}. \label{BF04}
\eea
Instead of using differential forms, we use the components of the $B$ and $F$ fields, we obtain
\bea
S= \frac{1}{2} \int d^4x \,\varepsilon^{\mu \nu \alpha \beta}\bigg (B_{IJ  \mu \nu}F^{IJ}_{\alpha \beta}-\frac{\beta}{2} B_{\mu \nu}^{IJ}B_{ \alpha \beta IJ}\bigg ),\label{BF05}
\eea
where $\varepsilon^{\mu\nu\alpha\beta}$ is the Levi-Civita symbol in $\mathcal M$. The Levi-Civita symbol is a totally antisymmetric quantity and, as a convention, we have $\varepsilon^{0123}=1$ and, in Minkowski space $\varepsilon_{0123}=-1$.

The equations of motion (EOM) are
\bea
0 &=&  \varepsilon^{\mu \nu \alpha \beta}(F_{\alpha \beta}^{IJ}-\beta B_{\alpha \beta}^{IJ} ), \label{06a} \\
0 &=& \varepsilon^{\mu \nu \alpha \beta}D_\nu B_{\alpha \beta}^{IJ},\label{06b}
\eea
where $D_\mu$ is the component of the covariant derivative $D\theta=d\theta+[A,\theta]$. For a $2-$form $\theta$, the explicit expression for the covariant derivative is
\bea
D_{\mu} \theta_{\alpha \beta}^{IJ}=\p_\mu \theta_{\alpha \beta}^{IJ} + A^{I}_{\mu K} \theta^{KJ}_{\alpha\beta}-A^{J}_{\mu K} \theta^{KI}_{\alpha\beta}.\label{07}
\eea
The EOM have a direct interpretation: Equation $(\ref{06a})$ states that the field strength and the background fields are parallel, while equation $(\ref{06b})$ states that the covariant derivative on the $B$ field, and as a consequence, the covariant derivative of the field strength is zero. Furthermore, on-shell, we can replace the background field $B_{\alpha \beta}^{IJ}=F_{\alpha \beta}^{IJ}/\beta$ and replace it on the $BF$ action $(\ref{BF05})$, obtaining a Yang-Mills (YM) like four-dimensional action. In \cite{Cattaneo}, the relation between the $BF$ model and the Yang-Mills was studied, while in \cite{Esc2}, the $BF$ model with a cosmological term gives exactly the YM theory for the $SU(N)$ group, showing that the $BF$ model can also be understood as a first-order action for the YM field.

\section{The Hamilton-Jacobi Analysis}

In order to perform a Hamilton-Jacobi analysis, we foliate the space-time $\mathcal M=R \times M_3$, being $M_3$ the space at constant time. The Lagrangian density from (\ref{BF05}) becomes
\bea
\mathcal L = \varepsilon^{ijk} B_{jkIJ}\p_0 A^{IJ}_i + \varepsilon^{ijk} A^{IJ}_0 D_i B_{jkIJ} + \varepsilon^{ijk} B_{0iIJ}( F^{IJ}_{jk}-\beta B_{jk}^{IJ}),\label{HJBF01}
\eea
where the lowercase Latin indices go from $1,2,3$ and denote the space coordinates, while the capital Latin indices represent the internal indices from the $SO(1,3)$ group. Furthermore, $\varepsilon^{ijk}\equiv \varepsilon^{0ijk}$ is the three-dimensional Levi-Civita symbol.

The canonical momenta $\pi_{\mu}^{IJ}$, $\Pi_{\mu \nu}^{IJ}$  conjugated to $A_{\mu}^{IJ}$ and $B_{\mu \nu}^{IJ}$  respectively, are defined by
\bea
\pi^{\mu}_{IJ} \equiv \frac{\p {\cal{L}}}{\p (\p_0 A_{\mu}^{IJ})},\quad \Pi^{\mu \nu}_{IJ} \equiv \frac{\p {\cal{L}}}{\p (\p_0 B_{\mu \nu}^{IJ})}.\label{HJBF02}
\eea
From the explicit expression of the Lagrangian, we notice that (\ref{HJBF01}) does not depend on any velocities $\p_0 A_\mu^{IJ}$, $\p_0 B_{\mu \nu}^{IJ}$. Therefore they are canonical constraints of the theory. For the other velocities, we have linear expressions that also represent constraints. Furthermore, the canonical Hamiltonian is given by
\bea
\mathcal{H}_0 = - \varepsilon^{ijk} A^{IJ}_0 D_i B_{jkIJ} - \varepsilon^{ijk} B_{0iIJ}( F^{IJ}_{jk}-\beta B_{jk}^{IJ}),\label{HJBF03}
\eea
and the canonical variables satisfy the following Poisson Brackets
\bea
\{A^{IJ}_\mu(x),\pi_{KL}^{\nu}(y)\} &=& \delta_{\mu}^\nu \Delta^{IJ}_{KL}\delta^3(x-y),  \label{HJBF04a} \\
\{B^{IJ}_{\mu \nu}(x), \Pi_{KL}^{\alpha \beta}(y)\}&=&\delta^{\mu\nu}_{\alpha\beta} \Delta^{IJ}_{KL}\delta^3(x-y),\label{HJBF04b}
\eea
where $\delta^{\mu\nu}_{\alpha\beta} =\frac{1}{2}(\delta^\mu_\alpha \delta^\nu_\beta-\delta^\mu_\beta \delta^\nu_\alpha)$ and $\Delta^{IJ}_{KL} \equiv \frac{1}{2}(\delta^I_K \delta^J_L-\delta^I_L \delta^J_K)$. The presence of these anti-symmetrised Kronecker deltas is the result of the antisymmetry of the indices of the gauge group.

According to the HJ formalism, we can define $\pi\equiv \p_0S$ where $S$ is the action. This definition allows us to write all the HJ PDE, or Hamiltonian, as
\bea
\mathcal{H}'&\equiv& \pi+{\cal{H}}=0, \label{HJBF05a} \\
{\cal{A}}^{0}_{IJ}&\equiv&\pi^0_{IJ}=0, \label{HJBF05b} \\
{\cal{A}}^{i}_{IJ}&\equiv&\pi^i_{IJ}-\varepsilon^{ijk} B_{jkIJ}=0, \label{HJBF05c} \\
{\cal{B}}^{\mu \nu}_{IJ}&\equiv&\Pi^{\mu \nu}_{IJ}=0. \label{HJBF05d}
\eea
The first Hamiltonian $\mathcal{H}'$ is associated with the time parameter $t\equiv x^0$. The Hamiltonians ${\cal{A}}^{\mu}_{IJ}$ are related to the momenta conjugated to the variable $A^{IJ}_\mu$ with correspondent parameters $\lambda^{IJ}_{\mu}$. Finally, the Hamiltonians ${\cal{B}}^{\mu}_{IJ}$, related to the momenta conjugated to the variable $B^{IJ}_{\mu\nu}$, have correspondent parameters $\omega^{IJ}_{\mu}$. These parameters play an important role in the definition of the fundamental differential
\bea
df=\int d^3y \bigg (\{f(x),{\cal{H}'}(y)\}dt+\{f(x),{\cal{A}}^{\mu}_{IJ}(y)\}d\lambda_\mu^{IJ}(y)+\{f(x),{\cal{B}}^{\mu \nu}_{IJ}(y)\}d\omega_{\mu \nu}^{IJ}(y) \bigg),\label{HJBF06}
\eea
and consequently, in the canonical structure of the theory.

The Hamiltonians that have vanishing Poisson brackets with themselves and all the remaining ones are called involutives. Otherwise, we have non-involutive Hamiltonians. From the set of HJ PDE above, we identity that $\cal{A}^{i}_{IJ}$ and $\cal{B}^{ij}_{IJ}$ are non-involutive, since
\bea
\{ \cal{A}^{i}_{IJ}(x),\cal{B}^{jk}_{KL}(y) \} = - \varepsilon^{ijk}\eta_{IR}\eta_{JS}\Delta^{RS}_{KL}\delta(x-y).\label{HJBF07}
\eea
with these Hamiltonians we can define generalized bracket. First, let us build the matrix between these constraints
\bea
M^{ijk}_{IJKL}(x,y)= \left(\begin{array}{cc}
0 & -\varepsilon^{ijk} \\
\varepsilon^{ijk} & 0 \\
\end{array}\right)\eta_{IR}\eta_{JS}\Delta^{RS}_{KL}\delta(x-y).\label{HJBF08}
\eea
This matrix has inverse, given by
\bea
(M^{-1})^{IJKL}_{ijk}(x,y)= \frac{1}{2}\left(\begin{array}{cc}
0 & -\varepsilon_{ijk} \\
\varepsilon_{ijk} & 0 \\
\end{array}\right)\eta^{IR}\eta^{JS}\Delta^{KL}_{RS}\delta(x-y).\label{HJBF09}
\eea
Once that this inverse exists, we can define the GB. Following equation (\ref{IC07}), we obtain the non-vanishing fundamental GB:
\bea
\{A_{iIJ}(x),B^{KL}_{kl}(y)\}^* &=& -\frac{1}{2}\varepsilon_{ikl}\Delta^{KL}_{IJ}\delta^3(x-y),  \label{HJBF10a}\\
\{A^{IJ}_\mu(x),\pi_{KL}^{\nu}(y)\}^*&=& \delta_{\mu}^\nu \Delta^{IJ}_{KL}\delta^3(x-y),\label{HJBF10b} \\
\{B^{IJ}_{0i}(x), \Pi_{KL}^{0j}(y)\}^*&=&\frac{1}{2}\delta^{i}_{j} \Delta^{IJ}_{KL}\delta^3(x-y).\label{HJBF10c}
\eea
Note that the PB (\ref{HJBF04a}) remains unaltered. Furthermore, from (\ref{HJBF10a}), we notice that $B^{IJ}_{ij}$ is now proportional to the canonical momenta of the variables $A^{IJ}_i$, in agreement with the Hamiltonian (\ref{HJBF05c}).

The GB redefine the dynamics of the system and the fundamental differential (\ref{HJBF06}) now takes the form:
\bea
df(x)=\int d^3y \bigg (\{f(x),{\cal{H}'}(y)\}^*dt+\{f(x),{\cal{A}}^{0}_{IJ}(y)\}^*d\lambda_0^{IJ}(y)+\{f(x),{\cal{B}}^{0k}_{IJ}(y)\}^*d\omega_{0 k}^{IJ}(y)\bigg).\label{HJBF11}
\eea
At this point we impose the IC for the remaining Hamiltonians: ${\cal{A}}^{0}_{IJ}(y) $ and ${\cal{B}}^{0k}_{IJ}(y)$. The condition $d{\cal{A}}^{0}_{IJ}(y)=0$ and $d{\cal{B}}^{0k}_{IJ}(y)=0$ introduce new Hamiltonians
\bea
{\mathcal C}_{IJ} &\equiv& \varepsilon^{ijk} D_i B_{jkIJ}, \label{HJBF12a}\\
{\cal{D}}^{i IJ} &\equiv& \varepsilon^{ijk}(F^{IJ}_{jk}-\beta B_{jk}^{IJ}),\label{HJBF12b}
\eea
note that these constraints can be identified as the EOM (\ref{06a}) and (\ref{06b}). Furthermore, the canonical Hamiltonian (\ref{HJBF03}) can now be written as the linear combination of the Hamiltonians:
\bea
{\cal{H}}_0=-A_0^{IJ}{\mathcal C}_{IJ}-B_{0k}^{IJ}{\cal{D}}^k_{IJ}.\label{HJBF13}
\eea
The IC are satisfied for the full set of Hamiltonians ${\cal{A}}^{0}_{IJ}$, ${\cal{B}}^{0i}_{IJ}$, ${\cal{C}}^{IJ}$, ${\cal{D}}^{i IJ}$. Moreover, they satisfy the following algebra:
\bea
\{{\cal{C}}^{IJ}(x),{\cal{C}}^{KL}(y)\}^*&=&\eta^{I L}{\cal{C}}^{JK}-\eta^{IK}{\cal{C}}^{JL}+\eta^{JK}{\cal{C}}^{IL}-\eta^{JL}{\cal{C}}^{IK}, \label{HJBF14a} \\
\{{\cal{C}}^{IJ}(x),{\cal{D}}^{KL}_k(y)\}^*&=& \eta^{IL}{\cal{D}}^{JK}_k-\eta^{IK}{\cal{D}}^{JL}_k+\eta^{JK}{\cal{D}}^{IL}_k-\eta^{JL}{\cal{D}}^{IK}_k,  \label{HJBF14b}                                                                                       \eea
and all other brackets between the Hamiltonians strictly zero. We conclude that the Hamiltonians are involutive.

\section{Characteristic equations}

The fundamental differential allows us to define the evolution of any function of the phase-space as a function of the time and the local parameters and it is built with the complete set of involutive Hamiltonians. Let us renamed them as
\bea
{\cal{H}}_{IJ} &\equiv& {\cal{A}}^{0}_{IJ} \to \lambda^{IJ} = \lambda^{IJ}_0, \nonumber \\
{\cal{H}}^{k}_{IJ} &\equiv& {\cal{B}}^{0k}_{IJ}\to \omega^{IJ}_k = \omega^{IJ}_{0k}, \nonumber \\
{\cal{G}}_{IJ} &\equiv& C_{IJ}\to \zeta^{IJ},\nonumber\\
{\cal{G}}^{k}_{IJ} &\equiv& {\cal{D}}^k_{IJ}\to \chi^{IJ}_ k. \nonumber
\eea
We have also introduced the parameters: $(\lambda^{IJ},\omega^{IJ}_k,\zeta^{IJ},\chi^{IJ}_ k)$ related to each Hamiltonian. Therefore, the fundamental differential is given by the linear combination
\bea
df(x) &=& \int d^3y\bigg (\{f(x),{\cal{H}'}(y)\}^*dt+\{f(x),{\cal{H}}_{IJ}(y)\}^*d\lambda^{IJ}
+\{f(x),{\cal{H}}^{k}_{IJ}(y)\}^*d\omega_k^{IJ}\nonumber\\
&+& \{f(x),{\cal{G}}^{k}_{IJ}\}^*d\chi^{IJ}_k+\{f(x), {\cal{G}}_{IJ}\}^* d\zeta^{IJ} \bigg).\label{CE01}
\eea

The characteristic equations are the ones that govern the evolution of the canonical variables of the phase-space. In our case, for the variables $A_\mu^{IJ}$ we have
\bea
dA_i^{IJ} &=& \left(D_i A^{IJ}_0+\beta B_{0i}^{IJ}\right)dt - D_i d\zeta^{IJ}-\beta d\chi^{IJ}_i  ,\label{CE02a}\\
dA_0^{IJ} &=& d\lambda^{IJ},\label{CE02b}
\eea
and for the $, B_{\mu\nu}^{IJ}$ field we have
\bea
dB_{ij}^{IJ} &=& \frac{1}{2} \varepsilon_{kij}\varepsilon^{kmn}\left(A_{0K}^{I}B_{mn}^{KJ}-A_{0K}^{J}B_{mn}^{KI}-2D_{m}^{x}B_{0n}^{IJ}\right)dt\nonumber\\
&+&-\left(B_{ijK}^{I}d\zeta^{KJ}-B_{ijK}^{J}d\zeta^{KI}\right)
-\left(D_{i}d\chi_{j}^{IJ}-D_{j}d\chi_{i}^{IJ}\right),\label{CE02c} \\
dB_{0i}^{IJ} &=& \frac{1}{2}d\omega_i^{IJ}.\label{CE02d}
\eea
Notice that the variables $A_0^{IJ}$ and $B_{0i}^{IJ}$ only depend on parameters $\lambda^{IJ}$ and $\omega^{IJ}_i$ respectively, enforcing that they only act as Lagrange multipliers. This can also be inferred from the form of the canonical Hamiltonian written in terms of the constraints.

The final form of the fundamental differential is given in terms of independent parameters, we can analyze the temporal evolution of the canonical variables independently. For the spatial components, we have
\bea
\p_0 A_i^{IJ}&=&D_i A^{IJ}_0+\beta  B_{0\mu}^{IJ},\label{CE03a}\\
\p_0 B_{ij}^{IJ}&=& \frac{1}{2} \epsilon_{kij}\varepsilon^{kmn}\left(A_{0K}^{I}B_{mn}^{KJ}-A_{0K}^{J}B_{mn}^{KI}-2D_{m}^{x}B_{0n}^{IJ}\right).\label{CE03b}
\eea
Equation (\ref{CE03a}) is in agreement with (\ref{06a}), while (\ref{CE03b}) becomes $\varepsilon^{kij}\left(D_{0}B_{\mu\nu}-2D_{j}B_{0k}\right)=0$, which is equivalent to the component $\mu=k$ of (\ref{06b}).

\subsection{Generators of canonical and gauge transformations}

The canonical transformations of the theory are obtained by setting $dt=0$ in the Characteristic Equations. Therefore, we have
\bea
\delta A_i^{IJ} &=& -D_i \delta\zeta^{IJ}-\beta \delta\chi^{IJ}_i  ,\label{GEN01a}\\
\delta A_0^{IJ} &=& \delta\lambda^{IJ},\label{GEN01b}\\
\delta B_{ij}^{IJ} &=& -\left(B_{ijK}^{I}\delta\zeta^{KJ}-B_{ijK}^{J}\delta\zeta^{KI}\right)
-\left(D_{i}\delta\chi_{j}^{IJ}-D_{j}\delta\chi_{i}^{IJ}\right),\label{GEN01c} \\
\delta B_{0i}^{IJ} &=& \frac{1}{2}\delta\omega_i^{IJ}.\label{GEN01d}
\eea
Then, the generator of canonical transformations is the linear combination of the involutive Hamiltonians:
\bea
G^{can} = \int d^3y\bigg ({\cal{H}}_{IJ}\delta\lambda^{IJ} + {\cal{H}}^{k}_{IJ}\delta \omega_k^{IJ} + {\cal{G}}^{k}_{IJ}\delta\chi^{IJ}_k + {\cal{G}}_{IJ} \delta\zeta^{IJ} \bigg),\label{GEN02}
\eea
once that
\bea
\delta A_{\mu}^{IJ}=\{A_{\mu}^{IJ},G^{can}\}^*\label{GEN03a} ,
\delta B_{\mu \nu}^{IJ}=\{B_{\mu \nu}^{IJ},G^{can}\}^*\label{GEN03b}.
\eea

In order to obtain the gauge transformations, i.e, the set of canonical transformations that leaves the Lagrangian (quasi-)invariant, we need to compute its variation $\delta {\cal{L}}$ induced by the field's canonical transformations and then impose $\delta {\cal{L}}=0$. This procedure generates constraints between the local parameters.

From (\ref{BF05}), we have the variation of Lagrangian
\bea
\delta\mathcal{L} &=& \varepsilon^{ijk}[(F^{IJ}_{jk}-\beta B^{IJ}_{jk})\delta B_{0iIJ}+(F^{IJ}_{0k}-\beta B^{IJ}_{0k})\delta B_{ijIJ}]\nonumber\\
&+&\varepsilon^{ijk}[2B_{0iIJ}D_{j}\delta A_{k}^{IJ}+B_{ijIJ}(D_{0}\delta A_{k}^{IJ}-D_{k}\delta A_{0}^{IJ})]\label{GEN04}
\eea
By replacing the expressions for $\delta B^{IJ}_{\mu\nu}$ and $\delta A^{IJ}_\mu$, we will have equations that will relate the four parameters: ($\lambda^{IJ},\omega^{IJ},\zeta^{IJ},\chi^{IJ}$). A good approach to solve $ \delta {\cal{L}} = 0$ is to consider special cases where some parameters are set to zero. From (\ref{GEN01a}) and (\ref{GEN01c}) we see that educated guesses are choosing $\delta\chi^{IJ}_i=0$ and $\delta\zeta^{IJ}=0$ independently.

First, let us set $\delta\zeta^{IJ}=0$. In this case, the equation for the variation $\delta \cal L=0$ becomes
\bea
0= \varepsilon^{ijk}\left[F_{ij}^{IJ}\left(\frac{1}{2}\delta\omega_{kIJ}+D_{0}\delta\chi_{kIJ}\right)-\beta B_{ij}^{IJ}\left(\frac{1}{2}\delta\omega_{kIJ}+D_{0}\delta\chi_{kIJ}+\frac{1}{\beta}D_{k}\delta\lambda_{IJ}\right)\right],
\label{GEN05}
\eea
which can be solved for $\frac{1}{2}\delta\omega_{kIJ}=-D_{0}\delta\chi_{kIJ}-\frac{1}{\beta}D_{k}\delta\lambda_{IJ}$. The term proportional to $B_{ij}^{IJ}$ becomes zero as well as the term proportional to $F_{ij}^{IJ}$ (up to a boundary term). Let us rename $\delta\lambda_{IJ}=-\beta \delta\chi_{0IJ}$. Therefore, we have
\bea
\delta A_\mu^{IJ} &=& -\beta \delta\chi^{IJ}_\mu  ,\label{GEN06a}\\
\delta B_{\mu\nu}^{IJ} &=& -\left(D_{\mu}\delta\chi_{\nu}^{IJ}-D_{\nu}\delta\chi_{\mu}^{IJ}\right),\label{GEN06b}
\eea
which is a shift translation for the $A$ field. Also notice that this translation only appears due to the cosmological term. Moreover, since we are dropping boundary terms this transformations leaves the Lagrangian quasi-invariant.

Now, as a second choice let us consider $\delta\chi^{IJ}_i=0$. Replacing in (\ref{GEN04}) we obtain
\bea
0=\varepsilon^{ijk}\left[\left(F_{ijIJ}-\beta B_{ijIJ}\right)\left(\frac{1}{2}\delta\omega_{k}^{IJ}+B_{0kK}^{I}\delta\zeta^{KJ}-B_{0kK}^{J}\delta\zeta^{KI}\right)
-B_{ijIJ}D_{k}\left(D_{0}\delta\zeta^{IJ}+\delta\lambda^{IJ}\right)\right].\label{GEN07}
\eea
This equation is zero for $\delta\lambda^{IJ}=- D_{0}\delta\zeta^{IJ}$ and $\delta\omega_{k}^{IJ}=-2\left(B_{0kK}^{I}\delta\zeta^{KJ}-B_{0kK}^{J}\delta\zeta^{KI}\right)$. This corresponds to the gauge transformation
\bea
\delta A_\mu^{IJ} &=& -D_\mu \delta\zeta^{IJ}  ,\label{GEN08a}\\
\delta B_{\mu\nu}^{IJ} &=& -\left(B_{\mu\nu K}^{I}\delta\zeta^{KJ}-B_{\mu\nu K}^{J}\delta\zeta^{KI}\right).\label{GEN08b}
\eea
This transformation does not depend on the parameter $\beta$, in contrast with (\ref{GEN06a}). Both, shift and gauge transformations obtained through the Hamilton-Jacobi formalism are in perfect agreement with the results presented in \cite{Mont}.

\section{Final remarks}

In this work  we have used the Hamilton-Jacobi formalism to analyze the $BF$ model with cosmological term. This procedure consisted in finding the complete set of Hamiltonians that generates the dynamical evolution of the system.  The Integrability Condition ensures that these hamiltonians are involutives. However, as we see from equation (\ref{HJBF07}), there are Hamiltonians which do not satisfy the IC. These non-involutive Hamiltonians can be eliminated if we define the generalized brackets which redefines the dynamic of the system, as shown in (\ref{HJBF10a}), (\ref{HJBF10b}) and (\ref{HJBF10c}). Related to each involutive Hamiltonians we have a independent parameter, one of them being the time.

An interesting feature of the Hamilton-Jacobi formalism is the fact that all the dynamics of the theory is given in terms of the fundamental differential. Particularly, for the four-dimensional $BF$ model the fundamental differential is given by (\ref{CE01}). Since the local parameters are linearly independent we can study the system's temporal evolution, from where we recover the equations of motion, as well as the canonical transformations (whenever we consider $dt=0$).

In order to obtain the symmetry transformations we vary the Lagrangian with respect to the fields, substituting its variations induced by the infinitesimal canonical transformations and set $\delta {\cal{L}}=0$. This procedure will generate relations between the local arbitrary parameters. Therefore, the generator of gauge transformations is equal to the generator of canonical transformation with dependent parameters. The Hamilton-Jacobi formalism provides a simple method to find the symmetries from the fundamental differential. The gauge and shift generators are obtained directly from the inner structure of the theory. This is an important motivation to use the Hamilton-Jacobi approach in the study of $BF$ theories.

Our plan is to extend this work to the Freidel-Starodubtsev $BF$ model, by adding a break invariance term. As shown in \cite{Horowitz} and \cite{Oda}, the study of the BF theory plus cosmological term give some of the properties of the $BF$ equivalents theories of gravity.

\section{Acknowledgements}
We would like to thank M. C. Bertin for reading the article and suggestions. GBG was supported by CNPq. BMP was partially supported by CNPq. CEV was supported by CNPq process 150407/2016-5. \\

\end{document}